\documentclass[iop]{emulateapj}

\usepackage{color}

\def\mr{\mathrm}
\def\ergsec{\mr{~erg~s}^{-1} }

\def\kms{\mr{~km\ s}^{-1} }

\def\msol{\mr{M}_{\odot}}
\def\lsol{\mr{L}_{\odot}}
\def\sfr{\ \msol\ \mr{yr}^{-1}}
\def\kmsmpc{\mr{\ km}\ \mr{s}^{-1}\ \mr{Mpc}^{-1}}

\def\cmc{\ \mr{cm^{-3}}}

\begin{document}

\title{Detection of Outflowing and Extraplanar Gas in Disks \\in an Assembling Galaxy Cluster at $z=0.37$}

\author{Emily Freeland\altaffilmark{1}}
\author{Kim-Vy H. Tran\altaffilmark{1,2}}
\author{Trevor Irwin\altaffilmark{1}}
\author{Lea Giordano\altaffilmark{2}}
\author{Am\'elie Saintonge\altaffilmark{3}}
\author{Anthony H. Gonzalez\altaffilmark{4}}
\author{Dennis Zaritsky\altaffilmark{5}}
\author{Dennis Just\altaffilmark{5}}


\altaffiltext{1}{George P. and Cynthia W. Mitchell Institute,
Department of Physics and Astronomy, Texas A\&M University, College
Station, TX 77843, freeland@physics.tamu.edu} 
\altaffiltext{2}{Institute for Theoretical Physics, University of
  Z\"urich, CH-8057 Z\"urich, Switzerland}
\altaffiltext{3}{Max-Planck-Institut f\"ur Extraterrestrische Physik,
Giessenbachstra\ss e, D-85748 Garching, Germany} 
\altaffiltext{4}{Department of Astronomy, University of Florida,
  Gainesville, FL 32611}
\altaffiltext{5}{Steward Observatory, University of Arizona, 933
N.~Cherry Avenue, Tucson, AZ 85721}

\begin{abstract}    We detect ionized gas characteristics indicative of winds
in three disk-dominated galaxies that are members of a super-group at $z=0.37$
that will merge to form a Coma-mass cluster.  All three galaxies are
IR-luminous ($L_{IR}>4\times10^{10}\ \lsol,~ \mathrm{SFR}>8\sfr$) and lie
outside the X-ray cores of the galaxy groups.  We find that the most
IR-luminous galaxy has strong blue and redshifted emission lines with
velocities of $\sim\pm200\kms$ and a third, blueshifted ($\sim 900 \kms$)
component.  This galaxy's line-widths (H$\beta$, [OIII]$\lambda$5007, [NII],
H$\alpha$) correspond to velocities of $100-1000\kms$.  We detect extraplanar
gas in two of three galaxies with SFR $> 8 \sfr$ whose orientations are
approximately edge-on and which have IFU spaxels off the stellar disk.  IFU
maps reveal that the extraplanar gas extends to $r_h\sim10$~kpc; [NII] and
H$\alpha$ line-widths correspond to velocities of $\sim200-400\kms$ in the disk
and decrease to $\sim50-150\kms$ above the disk.  Multi-wavelength observations
indicate that the emission is dominated by star formation.  Including the most
IR-luminous galaxy we find that $18\%$ of supergroup members with SFR $>8 \sfr$
show ionized gas characteristics indicative of outflows.  This is a lower limit
as showing that gas is outflowing in the remaining, moderately inclined,
galaxies requires a non-trivial decoupling of contributions to the emission
lines from rotational and turbulent motion.  Ionized gas mass loss in these
winds is $\sim 0.1\sfr$ for each galaxy, although the winds are likely to
entrain significantly larger amounts of mass in neutral and molecular gas.  

\end{abstract}


\section{Introduction}

Star formation can drive winds that transport gas to large distances from the
parent galaxy.  These outflows are nearly ubiquitous in infrared luminous
galaxies with star formation rates $>10\sfr$ \citep{2005ApJ...632..751R}.
In the local universe, extraplanar emission line gas is observed in most
galaxies with star formation rates $>5\sfr$ and in many cases there is evidence
that this gas is outflowing \citep{2005ARA&A..43..769V}.  Simulations
increasingly incorporate outflows in order to match a wide range of galaxy and
intergalactic gas observables \citep{2011arXiv1104.3156D,2011arXiv1103.3528D}.  

Starburst driven outflows are composed of multi-phase gas.  A tenuous, hot
($T\geq 10^6$ K) plasma created by stellar winds and supernovae ejecta is
thought to carry most of the energy and metals in the outflow
\citep{1985Natur.317...44C, 2007ApJ...658..258S} while neutral and molecular
interstellar gas clouds entrained in the hot outflow contribute the majority of
the mass \citep{2011ApJ...733L..16S, 2005ApJS..160..115R}.  Optical emission
lines are thought to be produced by turbulent mixing layers on the surface of
the cool entrained clouds as well as by diffuse ($T\sim 10^4$ K) ionized gas
\citep{2009ApJ...706.1571W}.

The fate of the ejected multi-phase gas is likely to depend on the galaxy mass
as well as environment.  For example, massive, isolated galaxies are able to
retain their hot, X-ray emitting gas reservoirs while low-mass galaxies are not
\citep{2010ApJ...715L...1M}.  Hot gas winds from galaxies in groups and
clusters may be confined by the dense intergalactic gas in these systems
\citep{2000ApJ...539..592B} or stripped away by it
\citep{2008ApJ...672L.103K,2006ApJ...647..910H}.  Gas which is stripped will
enrich the intergalactic medium with metals and entropy
\citep{2000MNRAS.315..689L,1999ApJ...511...34D}. 

Most galaxies in the local universe are in groups \citep{1983ApJS...52...61G,
2004MNRAS.348..866E} and groups are the building blocks of galaxy clusters
\citep{2008A&A...489...11B, 2009MNRAS.400..937M}.  As outflows are common among
star-forming galaxies, we expect to see many in group and field environments
whose star-forming galaxy populations are comparable at $0.3 < z < 0.55$
\citep{2011arXiv1107.2431T}.  We search for outflows from galaxies using
integral field unit (IFU) spectroscopic observations of SG1120-1202 (hereafter
SG1120), a system of four X-ray bright galaxy groups that will merge to form a
cluster comparable in mass to Coma \citep{2005ApJ...624L..73G}.  

We assume a flat cosmology with $H=71 \kmsmpc$, $\Omega_M=0.27$,
$\Omega_{vac}=0.73$.

\begin{deluxetable*}{lccccccc}
\tablecaption{Source Information\label{tab:sour}}
\tablehead{
\colhead{ID} & \colhead{$\alpha$} & \colhead{$\delta$} & \colhead{z} &
\colhead{ $M_*$\tablenotemark{a}}  & \colhead{$L_{1.4\ \mr{GHz}}$\tablenotemark{b}}   &
\colhead{ $L_{8-1000\mu m}$}        & \colhead{$SFR_{UV+IR}$} \\	
  & (J2000)   & (J2000)  &                  &($10^{11}\ \msol$) & ($10^{22}\ \mathrm{W\ Hz}^{-1}$)	& ($10^{11}\ \lsol$)	}
\startdata
S1	& 11 20 10.1	& -12 10 12	& 0.3692& 4.9 & 31 $\pm 0.2$	& 8.8\phantom{0}&150	\\  
S2	& 11 20 10.7    & -12 11 04	& 0.3501& 1.0 & 1.7$\pm 0.2$	& 0.54   	&9 	 \\
S3      & 11 20 17.7 	& -11 57 57	& 0.3557& 0.2 & -		& 0.50   	&11	
\enddata
\tablenotetext{a}{Details on the calculation of stellar masses, infrared luminosities and star formation rates can be found in \citet{2009ApJ...705..809T}.}
\tablenotetext{b}{The $5\sigma$ point source sensitivity is $1.1\times
10^{22}\ \mathrm{W\ Hz}^{-1}$.} 
\end{deluxetable*}

\section{Observations}

We used FLAMES/GIRAFFE \citep{2002Msngr.110....1P} on the VLT (PID:
082.B-0765) to take IFU spectroscopy of 60
SG1120 members in February 2009.  The instrument places 15 deployable
IFUs across the $25'$ field of view where each IFU is $3''\times 2''$
and made of 20 spaxels ($0.5 \arcsec$).  Using the LR08 setup, the
spectra cover a wavelength range of $822-940$ nm with spectral
resolution of 0.94\ \AA\ ($31 \kms$ for H$\alpha$ at $z=0.37$).  The
FWHM of the point spread function (PSF) ranged from $0.66-0.97''$.

We used the P3D software package \citep{2010A&A...515A..35S} to bias correct
the CCD images, trace the spectra on the chip, flat field, correct for
fiber-to-fiber transmission differences, and optimally extract the object
spectra using multi-profile deconvolution.  P3D also provides an error spectrum
with each extracted object spectrum.  We used IRAF to wavelength calibrate and
produce spectra interpolated to a dispersion of 0.2\ \AA\ $\mr{pix}^{-1}$.  The
sky background was subtracted using spectra from 12 dedicated sky fibers on
FLAMES.  The IFU spectra are reassembled spatially and emission lines are fit
with Gaussian components weighted by the associated error spectrum using custom
IDL and Supermongo programs.  The signal-to-noise (S/N) is calculated for each
line component in the fit and only components with S/N$>2$ are included in our
analysis.  We fit the central blended IFU spaxels in SG1120-S1 by hand with
IRAF/SPLOT using the three components in the [OIII]$\lambda$5007 line from the
multi-slit spectroscopy as a guide.

We have multi-slit spectroscopy with broader wavelength coverage of SG1120
\citep{2009ApJ...705..809T}.  The multi-slits are all $1''$ wide and the 2D
spectra summed over the brightest $1''$ along the slit to generate the 1D
spectra.  We measure systemic redshifts based on stellar absorption lines using
the IRAF/RVSAO package and fit the H$\beta$ and [OIII] emission lines by hand
with IRAF/SPLOT.

Key to our analysis is our extensive multi-wavelength dataset of SG1120 that
includes X-ray imaging from {\it Chandra}/ACIS \citep{2005ApJ...624L..73G},
total infrared luminosities from {\it Spitzer}/MIPS $24\mu$m imaging
\citep{2009ApJ...705..809T}, {\it Hubble}/ACS F814W imaging, GALEX UV imaging
\citep{2011arXiv1107.3838J}, and a 1.4 GHz radio continuum map from the VLA.
The X-ray, UV, mid-infrared, and radio observations are critical for
determining whether the velocity structure detected in the optical spectroscopy
is driven by star formation or by AGN.  

About a third of the SG1120 spectroscopically confirmed members (with $M_V <
-20.5$) are detected in {\it Spitzer} $24\mu$m imaging with star formation
rates $\geq3\sfr$ \citep{2009ApJ...705..809T}. Of these IR detected galaxies 17
have SFR $>8 \sfr$ ($L_{8-1000\mu m}> 4\times 10^{10}\ \lsol$, hereafter
$L_{IR}$).  We will discuss three of these galaxies, referred to arbitrarily as
S1, S2 and S3 in this paper. 

\begin{figure*} \epsscale{1.0} \plotone{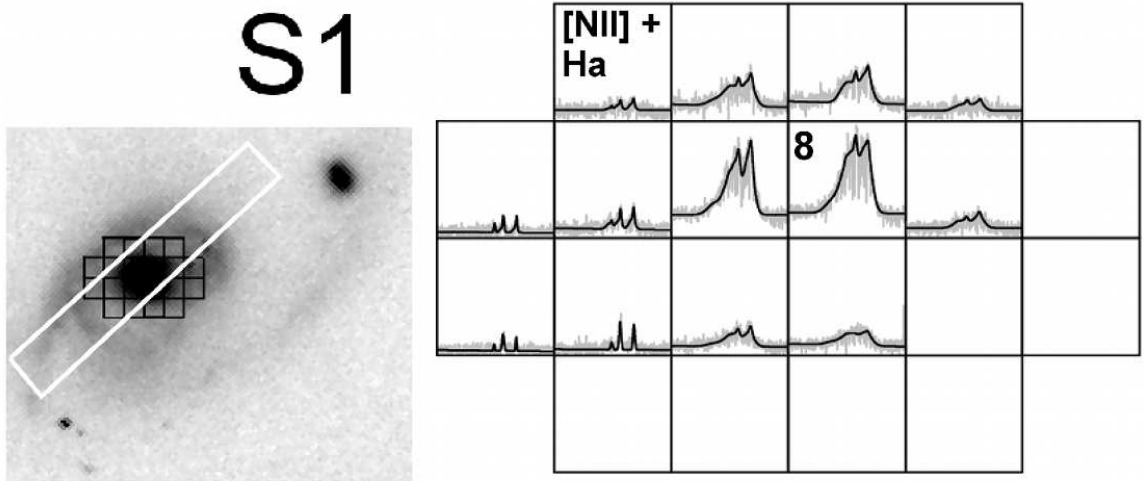} \caption{{\bf Left:} HST/ACS F814W image
with the FLAMES/GIRAFFE IFU ($3 \arcsec \times 2 \arcsec = 15\ \mathrm{kpc}
\times 10\ \mathrm{kpc}$) and single-slit ($1\arcsec$) footprints overlaid for source
SG1120-S1.  
{\bf Right:} The IFU data for S1 with spaxel 8 indicated.}\label{fig:s1}
\end{figure*}

\begin{figure*}\epsscale{1.0} \plotone{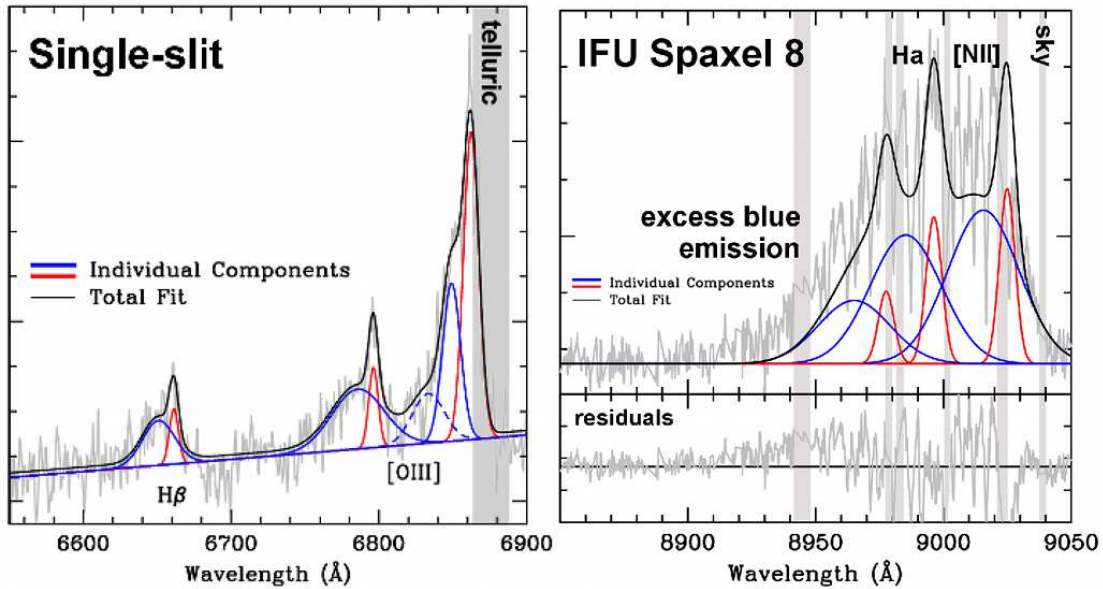} \caption{{\bf Left:} Single-slit
spectrum centered on the nucleus of S1 ($z=0.3692\pm 0.0002$).  The
line-of-sight velocity offsets are $\sim\pm200 \kms$ and FWHM line-widths are
$200 -1000 \kms$ for the main red and blue components.  The red side of the
[OIII]$\lambda$5007 line is affected by telluric absorption.  Forcing
consistency between the profile shapes for the H$\beta$ and [OIII]4959 lines
requires an additional broad component blueshifted by $900 \kms$ on the
[OIII]$\lambda$5007 line; this broad component is below our detection limit for
H$\beta$ and [OIII]$\lambda$4959.  The total fit is offset slightly high.  {\bf
Right:} Excess blue emission, seen after applying a two component model,
indicates that the second blueshifted component (seen on the
[OIII]$\lambda$5007 line in \ref{fig:s1}) is necessary.} \label{fig:1dandifu}
\end{figure*}

\section{Evidence of Outflowing and Extraplanar Gas in Group Galaxies}

\subsection{Luminous Infra-Red Galaxy: SG1120-S1}

SG1120-S1 is a massive super-group galaxy with an infrared luminosity of $9
\times 10^{11}\ \lsol$, i.e. nearly a ULIRG (Ultra-Luminous Infra-Red Galaxy).
Morphologically disturbed and connected to a nearby companion ($r_{proj}\sim25$
kpc) with a stellar tidal tail, SG1120-S1 is a poster child for merger-driven
star formation.  We find evidence of a strong gas outflow in SG1120-S1's
nucleus in both our IFU and single slit spectroscopy.  The $1''$-wide single
slit spectrum (Figs.~\ref{fig:s1} \& \ref{fig:1dandifu}) samples the galaxy's inner 5 kpc and
shows blue and redshifted components in H$\beta$ and [OIII] that differ from
the systemic velocity by $\sim\pm200\kms$.  By forcing the profile shapes of
the H$\beta$ and [OIII]4959 lines to be consistent, we find an additional broad
component in the [OIII]$\lambda$5007 line that is blueshifted by $900 \kms$;
note this broad component is much weaker than the other two components and thus
below our detection threshold for both H$\beta$ and [OIII]$\lambda$4959.  This
second blueshifted component is also necessary when fitting the blended spaxels
in the IFU data as the emission extends farther to the blue than would be
expected for only the single blueshifted component.  The line widths for
H$\beta$, [OIII]$\lambda$4959, and [OIII]$\lambda$5007 are also broad and range
from $\sim 1000 \kms$ for the blue component to $\sim200\kms$ for the red
component.  The combination of both blue and redshifted components in the
single-slit spectroscopic data indicate we may be viewing both sides of a
bipolar outflow, an expanding bubble, or an outflow plus infalling tidal
debris.

Our IFU spectroscopy maps the kinematics of the ionized gas as traced by [NII]
\& H$\alpha$ and reveals a complex of broad lines over multiple spaxels
(Fig.~\ref{fig:s1}).  There appears to be a single bright central spaxel that
is broadened into neighboring spaxels by the telescope PSF as indicated by the
unchanging red edge of the [NII]$\lambda$6583 line.  We show the IFU data for
illustration and use the single-slit data for all quantitative measurements of
outflow dynamics.  Beyond the galaxy's core ($r_{proj}>2.5$~kpc), the
[NII]-H$\alpha$ FWHM line-widths decrease and the velocity offsets are consistent
with rotation ($\pm 150 \kms$).  In spaxels with unblended lines we measure
line ratios ($\log$ [NII]/H$\alpha \sim 0$) that are best explained by shocked
gas as is commonly seen in the extranuclear regions of ULIRGS
\citep{2006ApJ...637..138M}.  Note that the true gas velocities are likely to
be higher because we are measuring only the motion along the line-of-sight.

Our multi-wavelength observations indicate that SG1120-S1 does not have a
significant AGN component (Table \ref{tab:sour}): our 1.4 GHz and $24\mu$m
measurements show that the galaxy falls on the infrared-radio relation for
local star-forming galaxies \citep{2001ApJ...554..803Y}.  SG1120-S1 is X-ray
detected but with an X-ray luminosity of $\sim 10^{42} \ergsec$, any embedded
AGN must be Compton thick \citep{2011arXiv1103.3212N}.  For comparison, most
U/LIRGs in this merger stage with similar $L_{IR}$ are starburst or composite
systems \citep{2010ApJ...709..884Y}. Star formation in U/LIRGs also tends to be
concentrated in the central kpc \citep{2001AJ....122.1213S,2000AJ....119..509S}
which corresponds to less than one spaxel in our IFU maps.  The lines in the
core spaxels are emitted primarily by gas that is participating in the outflow
and their line ratios appear to be dominated by shock excitation as is commonly
seen in starburst galaxies \citep{2010ApJ...711..818S}.  We conclude that
SG1120-S1 is not AGN dominated, thus the strong gas outflows detected in both
the IFU and single slit spectroscopy are driven primarily by star formation.

\begin{figure*} \epsscale{1.0} \plotone{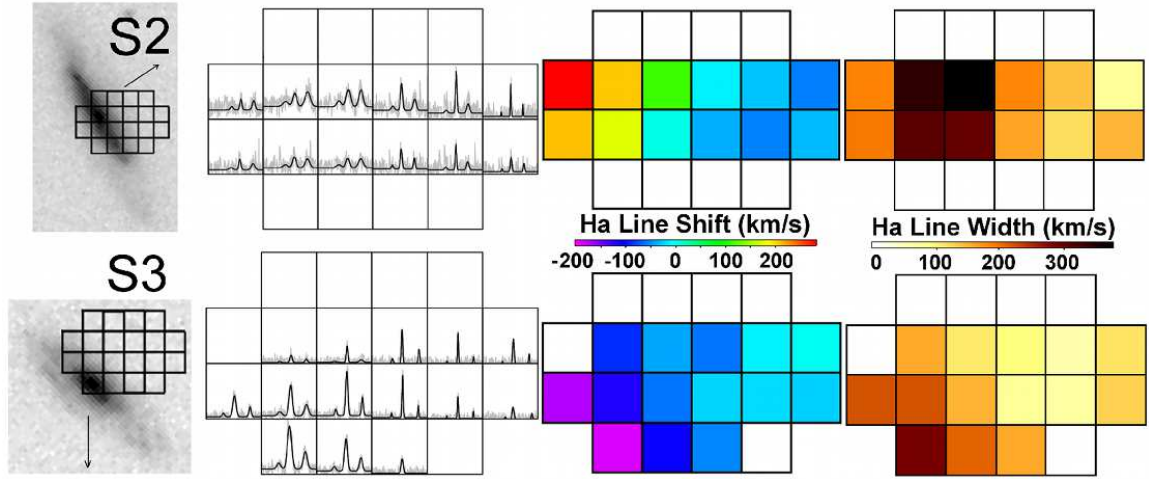} \caption{HST/ACS F814W image
with the FLAMES/GIRAFFE IFU footprint ($3 \arcsec \times 2 \arcsec = 15\
\mathrm{kpc} \times 10\ \mathrm{kpc}$) overlaid for sources SG1120-S2 and
SG1120-S3.  We center the IFU spaxel with the largest continuum value on the
nucleus of the galaxy.  The IFU footprint is populated with the spectra for
each spaxel centered on the [NII] \& H$\alpha$ lines.  The data are shown in
light grey and our fit to the emission lines in black.  Line shift and line
width maps are shown.  Errors on the line shifts (widths) are $68\kms$
($5-15\kms$) and $73 \kms$ ($5\kms$) for sources S2 and S3, respectively,
following from the $\sim 100 \kms$ errors on the systemic velocity.  Arrows
indicate the direction toward the nearest group center, G4 for S2 and G2 for
S3; group properties are detailed in \citet{2005ApJ...624L..73G} and
\citet{2009ApJ...705..809T}.  Both S2 and S3 have extraplanar gas to at
least $r_h\sim10$~kpc above their disks.  }\label{fig:s23} \end{figure*}

\subsection{Disk-dominated Members: SG1120-S2 \& SG1120-S3} 

In the local universe extraplanar gas is detected in most highly inclined
galaxies that have total infrared luminosities $>3\times 10^{10}\ \lsol$
\citep[$\mathrm{SFR}>5\sfr$;][]{2005ARA&A..43..769V}.  Our IFU observations
include three such super-group galaxies and we discover that two have
extraplanar emission with H$\alpha$ and [NII] FWHM line-widths of $50-150\kms$
(Fig.~\ref{fig:s23}).

SG1120-S2 is a disk-dominated member viewed nearly edge-on that lies on the
infrared-radio relation for local star-forming galaxies and is not detected
with {\it Chandra} (Table \ref{tab:sour}).  We detect [NII] and H$\alpha$
emission in the disk and, surprisingly, also at projected heights of
$r_{h}\sim7.5$~kpc above the disk (Fig.~\ref{fig:s23}).  In the spaxels
sampling the disk, the $\log$[NII]/H$\alpha$ ratios of 0 to 0.2 are consistent
with shocked gas and the measured line-widths correspond to gas velocities of
$\sim300-400\kms$.  In the extraplanar spaxels ($r_{h}>5$~kpc), the
$\log$[NII]/H$\alpha$ ratios of -0.2 to -0.4 are consistent with
photoionization by starlight and the line-widths correspond to gas velocities
of $\sim50-150~\kms$.  Our multi-wavelength observations indicate that as with
SG1120-S1, the shocked gas in the central region of SG1120-S2 is due to star
formation and not an AGN.

SG1120-S3 is also an inclined disk-dominated member with
comparable IR luminosity to SG1120-S2 (Table \ref{tab:sour}); it is not
detected in the radio nor X-ray observations.  The IFU maps show [NII] and
H$\alpha$ emission in both the disk and extraplanar spaxels
($r_{h}\sim10$~kpc), and the line-ratios are consistent with photoionization by
starlight.  The FWHM line-widths correspond to velocities of $\sim200-300~\kms$\ in
the disk spaxels and decrease to $\sim50-150~\kms$ above the disk.  With no
signs of an AGN, the gas motion is most likely driven by the ongoing star
formation.

In both group members where we detect extraplanar ionized gas, the emission
lines vary in terms of relative velocity and width from spaxel to spaxel
indicating that there is no PSF broadening in these sources.  As with
SG1120-S1, we measure only motion along the line-of-sight while the gas is
likely to be primarily moving perpendicular to the disk, i.e. the true gas
velocities are likely to be higher.  We cannot determine a net flow direction
for the extraplanar gas because the errors on the systemic velocity
($\sim100~\kms$) for these two galaxies are large compared to the velocity
shifts ($\sim 10-65 \kms$) in their extraplanar spaxels.  However, we do
confirm the existence of ionized gas at large scale heights above the disk of
both members.

\section{Outflow Gas Masses and Disk Gas Depletion Timescales}

To determine what happens to the gas in these three members, we first estimate
how much ionized gas is in the observed outflow.  For SG1120-S1, using the
H$\beta$ lines from our single-slit data and the relation in
\citet{2006MNRAS.370.1633H}\footnote{This estimate is subject to modulation by
a filling factor due to the clumpiness of the gas and assumes a spherical
outflow.}, we assume case B recombination \citep{1989agna.book.....O} and an
electron density\footnote{Electron densities of $50-300 \cmc$ are observed
between $1-10$ kpc for starburst superwinds \citep{1990ApJS...74..833H}.  We
cannot measure the electron density directly because the [SII] lines are
blended.} of $100 \cmc$ to estimate a total ionized gas mass of $M_{\mbox{\tiny
HII}}\sim 10^5\ \msol$ in the two components of the H$\beta$ line ($L_{H\beta}
= 2\times 10^{39}\ \ergsec$).  Next we estimate an outflow rate ($\dot{M}$) for
the ionized gas by comparing the mass inferred from the H$\beta$ emission to a
dynamical timescale.  Using the single-slit data we assume a radius of 1 kpc,
consistent with the extent of the emission lines, and an outflow velocity of
$900 \kms$ from the most blueshifted component on the [OIII]$\lambda$5007 line,
giving us $t_{dyn}=R/V\sim10^6$~yr.  For SG1120-S1, $\dot{M}\sim
M_H/t_{dyn}\approx 0.1~\sfr$ ; this is a lower limit because we cannot correct
for extinction.  


We stress that while the mass of ionized gas in the outflow is relatively
modest, the outflow is likely to contain larger amounts of neutral and
molecular gas.  We cannot directly trace with our observations the multi-phase
gas entrained in the outflow, but we note that SG1120-S1 has an intermediate $L_{IR}$ 
between the LIRG and low-z ULIRG samples in \citet{2005ApJS..160..115R} for
which typical outflows of neutral gas are, on average, $\sim17-42\sfr$.
SG1120-S1 has a similar star-formation rate and low AGN contribution to IRAS
13120-5453 for which $Herschel$ observations reveal a molecular gas outflow rate
of $130^{+390}_{-95}\ \sfr$ \citep{2011ApJ...733L..16S}.   

As SG1120-S1 loses gas through winds, it is also using up its reservoir of
dense molecular gas through strong star formation.  We can estimate the
molecular gas content of these galaxies using the $L_{FIR}-L_{CO}$ relation
\citep{1991ApJ...370..158S,2010MNRAS.407.2091G}.  SG1120-S1 has a molecular gas
reservoir of $\sim 9 \times 10^{10}\ \msol$ (using the Galactic conversion
factor).  We calculate a star formation rate for this galaxy of $170 \sfr$,
independent of its infrared emission, using its 1.4 GHz radio continuum
luminosity \citep{2003ApJ...586..794B}.  Assuming it maintains this star
formation rate and does not accrete new gas, SG1120-S1 will consume all of its
molecular gas in $\sim 500$ Myr.  If, as indicated above, SG1120-S1 is
expelling molecular gas at a rate comparable to its star formation rate then it
will run out of gas in half that time, or $250$ Myr.

For SG1120-S2 and SG1120-S3, we estimate the mass of ionized extraplanar gas
using the H$\alpha$ emission from our IFU observations
\citep{2006ApJ...650..693N}.  Adopting the same electron density as for
SG1120-S1 of $100~\cmc$, we estimate an ionized gas mass of $5\times10^6\
\msol$ for SG1120-S2 and $1\times 10^7\ \msol$ for SG1120-S3.  Assuming a
moderate outflow velocity of $\sim200~\kms$ consistent with winds from highly
inclined galaxies in the local universe \citep{1996ApJ...462..651L} the
dynamical timescale for the outflow to reach 10 kpc is $t_{dyn}\sim5 \times
10^7$ yr, thus the mass loss rate is $\dot{M}\sim0.1-0.2~\sfr$. Both galaxies
are likely to have $\sim 8\times 10^9\ \msol$ of molecular gas and will consume
all of this gas in the next $\sim800$~Myr if they maintain their current star
formation rates and do not accrete new gas.  

There are very few measurements of molecular outflows rates in galaxies which
are not ULIRGs.  In M82 \citep[$M_{tot}\sim 1\times 10^{10}\
\msol$;][]{1992ApJ...395..126S} whose $L_{IR}$ is similar to SG1120-S2 and
SG1120-S3, the molecular gas mass in the outflow is $\sim 30 \sfr$
\citep{2002ApJ...580L..21W}.  It is $1.6_{-1.2}^{+4.8} \sfr$ in NGC 253
\citep[$L_{IR}=2\times 10^{10}\ \lsol$, $M_*=4\times 10^{10}\
\msol$;][]{2011ApJ...736...24B,2011ApJ...733L..16S}.  Thus, SG1120-S2 and
SG1120-S3 may also be expelling significant quantities of molecular gas.

\section{Conclusions}

Using single-slit and IFU spectroscopy, we map the kinematics of ionized gas in
galaxies that are part of an assembling galaxy cluster at $z=0.37$.  Here we
present results on three IR-luminous, disk-dominated members that have ionized
gas characteristics indicative of outflows.

SG1120-S1 is the most IR-luminous super-group member ($L_{IR}=9\times10^{10}\
\lsol$, SFR of $150 \sfr$).  The H$\beta$ and [OIII]$\lambda4959,5007$ emission
lines in our single-slit data reveal gas that is blue and redshifted from the
systemic velocity by $\sim\pm200\kms$.  We find a third, weaker component in
the [OIII]$\lambda5007$ emission that is blueshifted by $900\kms$ but lies
below the noise in the [OIII]$\lambda4959$ and H$\beta$ lines.  The combination
of both blue and redshifted components in the single-slit spectroscopic data
indicate we are viewing both sides of a bipolar outflow or an expanding bubble, or
an outflow plus infalling tidal debris.

In the local universe extraplanar gas is detected in most highly inclined
galaxies that have star formation rates $>5\sfr$ \citep{2005ARA&A..43..769V}.
Our IFU maps target four highly inclined super-group members with SFR $> 5
\sfr$ of which two show extraplanar gas.  The IFU maps for SG1120-S2 and
SG1120-S3 reveal extraplanar [NII] and H$\alpha$ emission that is detected to
at least $\sim7.5-10$~kpc above the disk. 

Our radio, IR, and X-ray observations indicate that all three galaxies are
dominated by star formation.  If they maintain their current star formation
rates, and do not accrete new fuel, they will run out of fuel in
$\sim0.1-1$~Gyr. We estimate that the wind in each of the three galaxies drives
a relatively modest ionized gas mass flow of $\sim0.1\sfr$.  As discussed in
\S4, these galaxies may be ejecting molecular gas at a rate comparable to their
star formation rate.  

It remains unclear how future star formation in these galaxies will be affected
by these winds.  In simulations infalling galaxies continue to accrete gas for
$0.5-1$ Gyr after entering the larger halo \citep{2009MNRAS.399..650S} and a
number of observed characteristics of satellite galaxies can be explained by a
gradual reduction in star formation after infall \citep{2009MNRAS.394.1213W}.
A recent $GALEX$ study of the super-group S0 population finds no evidence for a
recent strong star formation episode which stopped abruptly in these galaxies
\citep{2011arXiv1107.3838J}.  If SG1120-2 and SG1120-3 are S0 progenitors whose
star formation will be quenched rapidly then they represent a new infalling
population.

About a third (31 galaxies) of the SG1120 spectroscopically confirmed members
(with $M_V < -20.5$) are detected in {\it Spitzer} $24\mu$m imaging with star
formation rates $\geq3\sfr$ ($L_{IR} > 2\times 10^{10}\ \lsol$)
\citep{2009ApJ...705..809T}.  Of these IR-detected galaxies 17 have SFR  $> 8
\sfr$.  Three are sources whose orientation is approximately edge-on and which
have IFU spaxels off the stellar disk and we detect extraplanar gas in two of
them, referred to here as S2 and S3.  Including S1 we find that $18\%$ of
supergroup members with SFR  $>8 \sfr$ show ionized gas characteristics
indicative of outflows.  This is a lower limit as showing that gas is
outflowing in the remaining, moderately inclined, galaxies requires a
non-trivial decoupling of contributions to the emission lines from rotational
and turbulent motion. Furthermore, $ 60\%$ of the IR-detected members have a
SFR $> 5 \sfr$ and, as a result, are likely to host a wind according to general
estimates of the frequency of outflows \citep{2005ARA&A..43..769V}.  

Further identification and study of galactic winds in groups and clusters is
needed to better understand their frequency and ability to transform galaxies
and intergalactic gas as a function of environment.

\acknowledgements 

KT acknowledges generous support from the Swiss National Science Foundation
(grant PP002-110576). This research has made use of the NASA/IPAC extragalactic
database (NED) which is operated by the Jet Propulsion Laboratory, Caltech,
under contract with the National Aeronautics and Space Administration.


\begin{thebibliography}

\bibitem[{{Bailin} {et~al.}(2011){Bailin}, {Bell}, {Chappell}, {Radburn-Smith},
  \& {de Jong}}]{2011ApJ...736...24B}
{Bailin}, J., {Bell}, E.~F., {Chappell}, S.~N., {Radburn-Smith}, D.~J., \& {de
  Jong}, R.~S. 2011, \apj, 736, 24

\bibitem[{{Bell}(2003)}]{2003ApJ...586..794B}
{Bell}, E.~F. 2003, \apj, 586, 794

\bibitem[{{Bou{\'e}} {et~al.}(2008){Bou{\'e}}, {Durret}, {Adami}, {Mamon},
  {Ilbert}, \& {Cayatte}}]{2008A&A...489...11B}
{Bou{\'e}}, G., {Durret}, F., {Adami}, C., {Mamon}, G.~A., {Ilbert}, O., \&
  {Cayatte}, V. 2008, \aap, 489, 11

\bibitem[{{Brown} \& {Bregman}(2000)}]{2000ApJ...539..592B}
{Brown}, B.~A. \& {Bregman}, J.~N. 2000, \apj, 539, 592

\bibitem[{{Chevalier} \& {Clegg}(1985)}]{1985Natur.317...44C}
{Chevalier}, R.~A. \& {Clegg}, A.~W. 1985, \nat, 317, 44

\bibitem[{{Dav{\'e}} {et~al.}(2011{\natexlab{a}}){Dav{\'e}}, {Finlator}, \&
  {Oppenheimer}}]{2011arXiv1104.3156D}
{Dav{\'e}}, R., {Finlator}, K., \& {Oppenheimer}, B.~D. 2011{\natexlab{a}},
  ArXiv e-prints

\bibitem[{{Dav{\'e}} {et~al.}(2011{\natexlab{b}}){Dav{\'e}}, {Oppenheimer}, \&
  {Finlator}}]{2011arXiv1103.3528D}
{Dav{\'e}}, R., {Oppenheimer}, B.~D., \& {Finlator}, K. 2011{\natexlab{b}},
  ArXiv e-prints

\bibitem[{{Davis} {et~al.}(1999){Davis}, {Mulchaey}, \&
  {Mushotzky}}]{1999ApJ...511...34D}
{Davis}, D.~S., {Mulchaey}, J.~S., \& {Mushotzky}, R.~F. 1999, \apj, 511, 34

\bibitem[{{Eke} {et~al.}(2004){Eke}, {Baugh}, {Cole}, {Frenk}, {Norberg},
  {Peacock}, {Baldry}, {Bland-Hawthorn}, {Bridges}, {Cannon}, {Colless},
  {Collins}, {Couch}, {Dalton}, {de Propris}, {Driver}, {Efstathiou}, {Ellis},
  {Glazebrook}, {Jackson}, {Lahav}, {Lewis}, {Lumsden}, {Maddox}, {Madgwick},
  {Peterson}, {Sutherland}, \& {Taylor}}]{2004MNRAS.348..866E}
{Eke}, V.~R., {Baugh}, C.~M., {Cole}, S. et al. 2004, \mnras, 348, 866

\bibitem[{{Geller} \& {Huchra}(1983)}]{1983ApJS...52...61G}
{Geller}, M.~J. \& {Huchra}, J.~P. 1983, \apjs, 52, 61

\bibitem[{{Genzel} {et~al.}(2010){Genzel}, {Tacconi}, {Gracia-Carpio},
  {Sternberg}, {Cooper}, {Shapiro}, {Bolatto}, {Bouch{\'e}}, {Bournaud},
  {Burkert}, {Combes}, {Comerford}, {Cox}, {Davis}, {Schreiber},
  {Garcia-Burillo}, {Lutz}, {Naab}, {Neri}, {Omont}, {Shapley}, \&
  {Weiner}}]{2010MNRAS.407.2091G}
{Genzel}, R., {Tacconi}, L.~J., {Gracia-Carpio}, J. et al. 2010, \mnras, 407, 2091

\bibitem[{{Gonzalez} {et~al.}(2005){Gonzalez}, {Tran}, {Conbere}, \&
  {Zaritsky}}]{2005ApJ...624L..73G}
{Gonzalez}, A.~H., {Tran}, K., {Conbere}, M.~N., \& {Zaritsky}, D. 2005, \apjl,
  624, L73

\bibitem[{{Heckman} {et~al.}(1990){Heckman}, {Armus}, \&
  {Miley}}]{1990ApJS...74..833H}
{Heckman}, T.~M., {Armus}, L., \& {Miley}, G.~K. 1990, \apjs, 74, 833

\bibitem[{{Hester}(2006)}]{2006ApJ...647..910H}
{Hester}, J.~A. 2006, \apj, 647, 910

\bibitem[{{Holt} {et~al.}(2006){Holt}, {Tadhunter}, {Morganti}, {Bellamy},
  {Gonz{\'a}lez Delgado}, {Tzioumis}, \& {Inskip}}]{2006MNRAS.370.1633H}
{Holt}, J., {Tadhunter}, C., {Morganti}, R., {Bellamy}, M., {Gonz{\'a}lez
  Delgado}, R.~M., {Tzioumis}, A., \& {Inskip}, K.~J. 2006, \mnras, 370, 1633

\bibitem[{{Just} {et~al.}(2011){Just}, {Zaritsky}, {Tran}, {Gonzalez},
  {Kautsch}, \& {Moustakas}}]{2011arXiv1107.3838J}
{Just}, D.~W., {Zaritsky}, D., {Tran}, K.-V.~H., {Gonzalez}, A.~H., {Kautsch},
  S.~J., \& {Moustakas}, J. 2011, ArXiv e-prints

\bibitem[{{Kawata} \& {Mulchaey}(2008)}]{2008ApJ...672L.103K}
{Kawata}, D. \& {Mulchaey}, J.~S. 2008, \apjl, 672, L103

\bibitem[{{Lehnert} \& {Heckman}(1996)}]{1996ApJ...462..651L}
{Lehnert}, M.~D. \& {Heckman}, T.~M. 1996, \apj, 462, 651

\bibitem[{{Lloyd-Davies} {et~al.}(2000){Lloyd-Davies}, {Ponman}, \&
  {Cannon}}]{2000MNRAS.315..689L}
{Lloyd-Davies}, E.~J., {Ponman}, T.~J., \& {Cannon}, D.~B. 2000, \mnras, 315,
  689

\bibitem[{{McGee} {et~al.}(2009){McGee}, {Balogh}, {Bower}, {Font}, \&
  {McCarthy}}]{2009MNRAS.400..937M}
{McGee}, S.~L., {Balogh}, M.~L., {Bower}, R.~G., {Font}, A.~S., \& {McCarthy},
  I.~G. 2009, \mnras, 400, 937

\bibitem[{{Monreal-Ibero} {et~al.}(2006){Monreal-Ibero}, {Arribas}, \&
  {Colina}}]{2006ApJ...637..138M}
{Monreal-Ibero}, A., {Arribas}, S., \& {Colina}, L. 2006, \apj, 637, 138

\bibitem[{{Mulchaey} \& {Jeltema}(2010)}]{2010ApJ...715L...1M}
{Mulchaey}, J.~S. \& {Jeltema}, T.~E. 2010, \apjl, 715, L1

\bibitem[{{Nardini} \& {Risaliti}(2011)}]{2011arXiv1103.3212N}
{Nardini}, E. \& {Risaliti}, G. 2011, ArXiv e-prints

\bibitem[{{Nesvadba} {et~al.}(2006){Nesvadba}, {Lehnert}, {Eisenhauer},
  {Gilbert}, {Tecza}, \& {Abuter}}]{2006ApJ...650..693N}
{Nesvadba}, N.~P.~H., {Lehnert}, M.~D., {Eisenhauer}, F., {Gilbert}, A.,
  {Tecza}, M., \& {Abuter}, R. 2006, \apj, 650, 693

\bibitem[{{Osterbrock}(1989)}]{1989agna.book.....O}
{Osterbrock}, D.~E. 1989, {Astrophysics of gaseous nebulae and active galactic
  nuclei}, ed. {Osterbrock, D.~E.}

\bibitem[{{Pasquini} {et~al.}(2002){Pasquini}, {Avila}, {Blecha}, {Cacciari},
  {Cayatte}, {Colless}, {Damiani}, {de Propris}, {Dekker}, {di Marcantonio},
  {Farrell}, {Gillingham}, {Guinouard}, {Hammer}, {Kaufer}, {Hill}, {Marteaud},
  {Modigliani}, {Mulas}, {North}, {Popovic}, {Rossetti}, {Royer}, {Santin},
  {Schmutzer}, {Simond}, {Vola}, {Waller}, \& {Zoccali}}]{2002Msngr.110....1P}
{Pasquini}, L., {Avila}, G., {Blecha}, A. et al. 2002, The Messenger, 110, 1

\bibitem[{{Rupke} {et~al.}(2005{\natexlab{a}}){Rupke}, {Veilleux}, \&
  {Sanders}}]{2005ApJ...632..751R}
{Rupke}, D.~S., {Veilleux}, S., \& {Sanders}, D.~B. 2005{\natexlab{a}}, \apj,
  632, 751

\bibitem[{{Rupke} {et~al.}(2005{\natexlab{b}}){Rupke}, {Veilleux}, \&
  {Sanders}}]{2005ApJS..160..115R}
---. 2005{\natexlab{b}}, \apjs, 160, 115

\bibitem[{{Sanders} {et~al.}(1991){Sanders}, {Scoville}, \&
  {Soifer}}]{1991ApJ...370..158S}
{Sanders}, D.~B., {Scoville}, N.~Z., \& {Soifer}, B.~T. 1991, \apj, 370, 158

\bibitem[{{Sandin} {et~al.}(2010){Sandin}, {Becker}, {Roth}, {Gerssen},
  {Monreal-Ibero}, {B{\"o}hm}, \& {Weilbacher}}]{2010A&A...515A..35S}
{Sandin}, C., {Becker}, T., {Roth}, M.~M., {Gerssen}, J., {Monreal-Ibero}, A.,
  {B{\"o}hm}, P., \& {Weilbacher}, P. 2010, \aap, 515, A35+

\bibitem[{{Sharp} \& {Bland-Hawthorn}(2010)}]{2010ApJ...711..818S}
{Sharp}, R.~G. \& {Bland-Hawthorn}, J. 2010, \apj, 711, 818

\bibitem[{{Simha} {et~al.}(2009){Simha}, {Weinberg}, {Dav{\'e}}, {Gnedin},
  {Katz}, \& {Kere{\v s}}}]{2009MNRAS.399..650S}
{Simha}, V., {Weinberg}, D.~H., {Dav{\'e}}, R., {Gnedin}, O.~Y., {Katz}, N., \&
  {Kere{\v s}}, D. 2009, \mnras, 399, 650

\bibitem[{{Sofue} {et~al.}(1992){Sofue}, {Reuter}, {Krause}, {Wielebinski}, \&
  {Nakai}}]{1992ApJ...395..126S}
{Sofue}, Y., {Reuter}, H.-P., {Krause}, M., {Wielebinski}, R., \& {Nakai}, N.
  1992, \apj, 395, 126

\bibitem[{{Soifer} {et~al.}(2000){Soifer}, {Neugebauer}, {Matthews}, {Egami},
  {Becklin}, {Weinberger}, {Ressler}, {Werner}, {Evans}, {Scoville}, {Surace},
  \& {Condon}}]{2000AJ....119..509S}
{Soifer}, B.~T., {Neugebauer}, G., {Matthews}, K. et al. 2000, \aj, 119, 509

\bibitem[{{Soifer} {et~al.}(2001){Soifer}, {Neugebauer}, {Matthews}, {Egami},
  {Weinberger}, {Ressler}, {Scoville}, {Stolovy}, {Condon}, \&
  {Becklin}}]{2001AJ....122.1213S}
{Soifer}, B.~T., {Neugebauer}, G., {Matthews}, K. et al. 2001, \aj, 122, 1213

\bibitem[{{Strickland} \& {Heckman}(2007)}]{2007ApJ...658..258S}
{Strickland}, D.~K. \& {Heckman}, T.~M. 2007, \apj, 658, 258

\bibitem[{{Sturm} {et~al.}(2011){Sturm}, {Gonz{\'a}lez-Alfonso}, {Veilleux},
  {Fischer}, {Graci{\'a}-Carpio}, {Hailey-Dunsheath}, {Contursi}, {Poglitsch},
  {Sternberg}, {Davies}, {Genzel}, {Lutz}, {Tacconi}, {Verma}, {Maiolino}, \&
  {de Jong}}]{2011ApJ...733L..16S}
{Sturm}, E., {Gonz{\'a}lez-Alfonso}, E., {Veilleux}, S. et al. 2011, \apjl, 733, L16+

\bibitem[{{Tran} {et~al.}(2009){Tran}, {Saintonge}, {Moustakas}, {Bai},
  {Gonzalez}, {Holden}, {Zaritsky}, \& {Kautsch}}]{2009ApJ...705..809T}
{Tran}, K., {Saintonge}, A., {Moustakas}, J., {Bai}, L., {Gonzalez}, A.~H.,
  {Holden}, B.~P., {Zaritsky}, D., \& {Kautsch}, S.~J. 2009, \apj, 705, 809

\bibitem[{{Tyler} {et~al.}(2011){Tyler}, {Rieke}, {Wilman}, {McGee}, {Bower},
  {Bai}, {Mulchaey}, {Parker}, {Shi}, \& {Pierini}}]{2011arXiv1107.2431T}
{Tyler}, K., {Rieke}, G.~H., {Wilman}, D.~J. et al. 2011, ArXiv e-prints

\bibitem[{{Veilleux} {et~al.}(2005){Veilleux}, {Cecil}, \&
  {Bland-Hawthorn}}]{2005ARA&A..43..769V}
{Veilleux}, S., {Cecil}, G., \& {Bland-Hawthorn}, J. 2005, \araa, 43, 769

\bibitem[{{Walter} {et~al.}(2002){Walter}, {Weiss}, \&
  {Scoville}}]{2002ApJ...580L..21W}
{Walter}, F., {Weiss}, A., \& {Scoville}, N. 2002, \apjl, 580, L21

\bibitem[{{Weinmann} {et~al.}(2009){Weinmann}, {Kauffmann}, {van den Bosch},
  {Pasquali}, {McIntosh}, {Mo}, {Yang}, \& {Guo}}]{2009MNRAS.394.1213W}
{Weinmann}, S.~M., {Kauffmann}, G., {van den Bosch}, F.~C., {Pasquali}, A.,
  {McIntosh}, D.~H., {Mo}, H., {Yang}, X., \& {Guo}, Y. 2009, \mnras, 394, 1213

\bibitem[{{Westmoquette} {et~al.}(2009){Westmoquette}, {Gallagher}, {Smith},
  {Trancho}, {Bastian}, \& {Konstantopoulos}}]{2009ApJ...706.1571W}
{Westmoquette}, M.~S., {Gallagher}, J.~S., {Smith}, L.~J., {Trancho}, G.,
  {Bastian}, N., \& {Konstantopoulos}, I.~S. 2009, \apj, 706, 1571

\bibitem[{{Yuan} {et~al.}(2010){Yuan}, {Kewley}, \&
  {Sanders}}]{2010ApJ...709..884Y}
{Yuan}, T., {Kewley}, L.~J., \& {Sanders}, D.~B. 2010, \apj, 709, 884

\bibitem[{{Yun} {et~al.}(2001){Yun}, {Reddy}, \&
  {Condon}}]{2001ApJ...554..803Y}
{Yun}, M.~S., {Reddy}, N.~A., \& {Condon}, J.~J. 2001, \apj, 554, 803

\end{thebibliography}
\end{document}